# Nature of Structural Changes Near the Magnetic Ordering Temperature in Small-Ion Rare Earth Perovskites $RMnO_3$

*T. Yu[1], T. A. Tyson[1,*], H. Y. Chen[1], A. M. M. Abeykoon[2], Y.-S. Chen[3], and K. H. Ahn[1]*

[1]Department of Physics, New Jersey Institute of Technology, Newark, NJ 07102
[2]National Synchrotron Light Source, Brookhaven National Laboratory, Upton, NY 11973
[3]ChemMatCARS, University of Chicago and Advanced Photon Source, Argonne National Laboratory, IL 60439

Detailed structural measurements were conducted on a new perovskite, $ScMnO_3$, and on orthorhombic $LuMnO_3$. Complementary density functional theory (DFT) calculations were carried out, and predict that $ScMnO_3$ possesses E-phase magnetic order at low temperature with displacements of the Mn sites (relative to the high temperature state) of ~0.07 Å, compared to ~ 0.04 Å predicted for $LuMnO_3$. However, detailed local, intermediate and long-range structural measurements by x-ray pair distribution function analysis, single crystal x-ray diffraction and x-ray absorption spectroscopy, find no local or long- range distortions on crossing into the low temperature E-phase of the magnetically ordered state. The measurements place upper limits on any structural changes to be at most one order of magnitude lower than density functional theory predictions and suggest that this theoretical approach does not properly account for the spin-lattice coupling in these oxides and may possibly predict the incorrect magnetic order at low temperatures. The results suggest that the electronic contribution to the electrical polarization dominates and should be properly treated in theoretical models.

## I. INTRODUCTION

Depending on the ionic radius of the rare-earth elements, $RMnO_3$ (*R*: Rare-earth) manganites naturally form two types of structures: orthorhombic (*R* = La-Dy) and hexagonal (*R* = Sc, Y, Ho-Lu) structures[1]. Materials with both hexagonal and orthorhombic structures are found to exhibit simultaneous ferroelectric and magnetic properties (multiferroic) and have potential in novel spintronic applications. Using high temperature and high pressure syntheses methods[2], film deposition techniques[3] and chemistry solution methods[4], the hexagonal structure can be stabilized as an orthorhombic phase[5]. The orthorhombic perovskite materials with small ion size (such as $HoMnO_3$ and $LuMnO_3$) have become the focus of detailed studies because of the possible existence of mainly electronically-driven ferroelectricity and are a topic of intense research (see refs. in [6]). High electric polarization values approaching of the values measured in hexagonal phase systems such as $YMnO_3$ are predicted by density functional theory (DFT). The models also predict large atomic displacements form the centrosymmetric state.

Metastable o-$RMnO_3$ (orthorhombic perovskite structure) is difficult to prepare in a single crystalline form. Hence most studies have been conducted on polycrystalline powders. The polycrystalline o-$HoMnO_3$ exhibits a dielectric constant which increases below $T_N$ and a spontaneous polarization below $T_L$[7]. Because of $Ho^{3+}$ spin ordering at low temperature, the polarization shows changes that strongly depend on magnetic field and temperature. Large spontaneous polarization along the *a*-axis was suggested for E-type o-$RMnO_3$ in a model that includes atomic and electronic degrees of freedom, the displacement mechanism, and spin-orbit coupling[8]. For o-$RMnO_3$ systems, recent measurements of polarization of o-$RMnO_3$ yield values of 0.025 $\mu C/cm^2$ at 4.5 K and 0 T for $YMnO_3$, 0.008 $\mu C/m^2$ at 4.5 K and 0 T for $HoMnO_3$[7], ~ 0.2000 $\mu C/cm^2$ at 10 K for $DyMnO_3$[5] and 0.080 $\mu C/cm^2$ at 10K for $TbMnO_3$[9]. Measurements on the phase diagram of perovskite Lu doped and $YMnO_3$ yield a bulk polarization value of ~0.4000 $\mu C/cm^2$ at ~5 K significantly below the magnetic ordering temperature of ~40K[10]. These values are only modest compared with the value of 75.0$\mu C/cm^2$ for a typical ferroelectric material $PbTiO_3$, and the ferroelectricity of which is related to off-center displacements of Ti by ~0.2 Å.

Although coupling of magnetization and polarization is known to exist in o-$RMnO_3$, probing the properties in multiple length scales will provide a clearer picture of the nature of the electric polarization by specifying its true origin (electronic and/or atomic) and possibly provide paths to enable enhancement and coupling with magnetic fields. Few experimental studies have explored the changes in structure near the magnetic transition temperature. Single crystal diffraction measurements on orthorhombic $YMnO_3$ reveal a reduction in space group symmetry from Pbam to $P2_1nm$ but with atomic displacements of the order $10^{-3}$ Å[11] on crossing from above (50 K) to below (21 K) the magnetic transition temperature. It was found that the DFT predictions for Mn and O displacements in the E phase relative the paramagnetic phase were reversed compared to what was observed experimentally, which indicates serious inconsistency between DFT and experimental results. In addition to this observation, more advanced hybrid density functional models applied to orthorhombic $HoMnO_3$ predict a value of P ~ 2 $\mu$C/cm$^2$ compared to P ~ 6 $\mu$C/cm$^2$ by standard DFT methods[12].

* E-mail: : tyson@njit.edu

No systematic structural study of the E-phase perovskites has been conducted covering both long range and short range structure and compared them directly with DFT calculations on simple E phase systems with only Mn magnetic sties. In the current work, perovskite $ScMnO_3$ was prepared and detailed temperature-dependent X-ray absorption measurements (XAFS), pair distribution function measurements (PDF) and single crystal diffractions measurements were done extending down to temperatures near 10 K. DFT modeling was conducted for comparison and predicted large displacement of the Mn ions (twice the predicted value for the $LuMnO_3$ system). The DFT models predict that this system possesses E-phase magnetic order. However, the measurements reveal no significant local or long-range distortions and contradict the predicted results of DFT models, revealing that DFT does not properly account for the spin- lattice coupling in these oxides and possibly predicts the incorrect magnetic order at low temperatures. Similar structural measurements and modeling were conducted on perovskite $LuMnO_3$, yielding the same conclusions.

## II. EXPERIMENTAL AND MODELING METHODS

Single crystal samples of perovskite $ScMnO_3$ and LuMnO3 were prepared as reported in our previous works[13, 14]. Single crystal diffraction measurements were carried out on $ScMnO_3$ at the beamline 15-ID-B of the Advanced Photon Source at Argonne National Laboratory with a wavelength of 0.41328 Å. Refinement of the single crystal data was conducted using the program SHELXL[15] after the reflections were corrected for absorption (see Ref. [12]). For pair distribution function (PDF) measurements on the $ScMnO_3$, powder samples were ground to 500 mesh size. Experiments were conducted at the beamline X17A at the National Synchrotron Light Source (NSLS) at Brookhaven National Laboratory. The wavelength was set at 0.18390 Å and the measurements were conducted using a Perkin Elmer detector with the sample-to-detector distance of 204 mm. For the large angle measurement, a maximum scattering vector magnitude, $Q_{max}$ = 25 Å$^{-1}$, was used in data reduction. For fits in r-space varying ranges of 1.5 < r < $r_{max}$ [$r_{max}$= 5 Å (local structure), 15 Å (short range structure) or 40 Å (intermediate range structure)] were used. PDF data reduction was conducted by the methods in Ref. [16]. The PDF measurements for the $LuMnO_3$ system were conducted at NSLS beamline X17B3 at an x-ray of a wavelength of 0.150053 Å, with data reduction carried out up to $Q_{max}$ = 27.5 Å$^{-1}$. For XAFS measurements, polycrystalline samples were prepared by grinding and sieving the materials (500 mesh) and brushing them onto Kapton tape. Layers of tape were stacked to produce a uniform sample for transmission measurements with jump $\mu$t ~1, where1/$\mu$ is the absorption length. Spectra for $ScMnO_3$ and $LuMnO_3$ were measured at the NSLS beamlines X19A (Sc K-edge, 4492 eV), X11A (Lu $L_3$-Edge, 9244 eV) and X3A (Mn K-edge, 6539 eV) at Brookhaven National Laboratory. Measurements were made on warming from 30 K to 300 K with the sample attached to the cold finger of a cryostat. Three to four scans were taken at each temperature. The uncertainty in temperature is less than 0.2 K. At the Mn K-Edge, a Mn foil reference was employed for energy calibration. The reduction of the X-ray absorption fine-structure (XAFS) data was performed using standard procedures[17]. For multi-shell fits at the Mn K-Edge, the k-range, $\sqrt{(E-E_0)2m/\hbar}$ , 1.56 Å$^{-1}$ < k < 12.53 Å$^{-1}$ and the R-range 0.71 Å < R < 3.64 Å were used with $S_0^2$ = 0.90. For the Sc K-edge, the k-range 2.63 < k < 12.65 Å$^{-1}$ and the R-range 0.95 < R < 3.62 Å were used with $S_0^2$ = 0.90. Coordination numbers for the atomic shells were fixed to the crystallographic values. For the $LuMnO_3$ $L_3$-edge, multi-shell fits were conducted for the k-range 2.75 Å$^{-1}$ < k < 16.1 Å$^{-1}$ and the R-range 1.07 Å < R < 4.32 Å were used with $S_0^2$ = 1.1. For heat capacity and magnetic measurements, a Quantum Design Physical Properties Measurements System was utilized.

The same approach taken by Picozzi *et al.*[18] with the VASP code[19] was implemented in our DFT simulations on $LuMnO_3$ and $ScMnO_3$ . All of our DFT structural optimizations were conducted using 40 atoms cells to reflect the magnetic order for the *E*-phase and treat all magnetic orderings in the same manner. Simulations were carried out with the experimentally derived lattice parameters, but the atomic positions were optimized to reduce the forces on the atoms to less than 0.003 eV/Å. Total energies were thus found for each spin configuration.

## III. RESULTS AND DISCUSSION

### A. Structure from DFT Simulations

Figure 1 shows the full structure of the perovskite $ScMnO_3$ system with two distinct Mn sites and three unique O sites labeled (see Ref. 12). The deviation from standard orthorhombic phase is small with β = 93.6 at room temperature. In this structure, the Mn-O bonds in the $MnO_6$ polyhedra are 1.920 Å, 1.967 Å and 2.138 Å at the Mn1 site and are 1.902 Å, 1.930 Å and 2.320 Å at the Mn2 site. The average spread in bond distances (standard deviation) is above that of the $LuMnO_3$ system. A unique R (Sc) site exists as in the case of $LuMnO_3$ and the classic orthorhombic systems. It is found[12] that the pre-edge 1s-to-3d (3d hybridized-with-p) feature in the near edge is better resolved in $ScMnO_3$ compared to $LuMnO_3$, indicating more localized 3d states in the perovskite $ScMnO_3$ system.

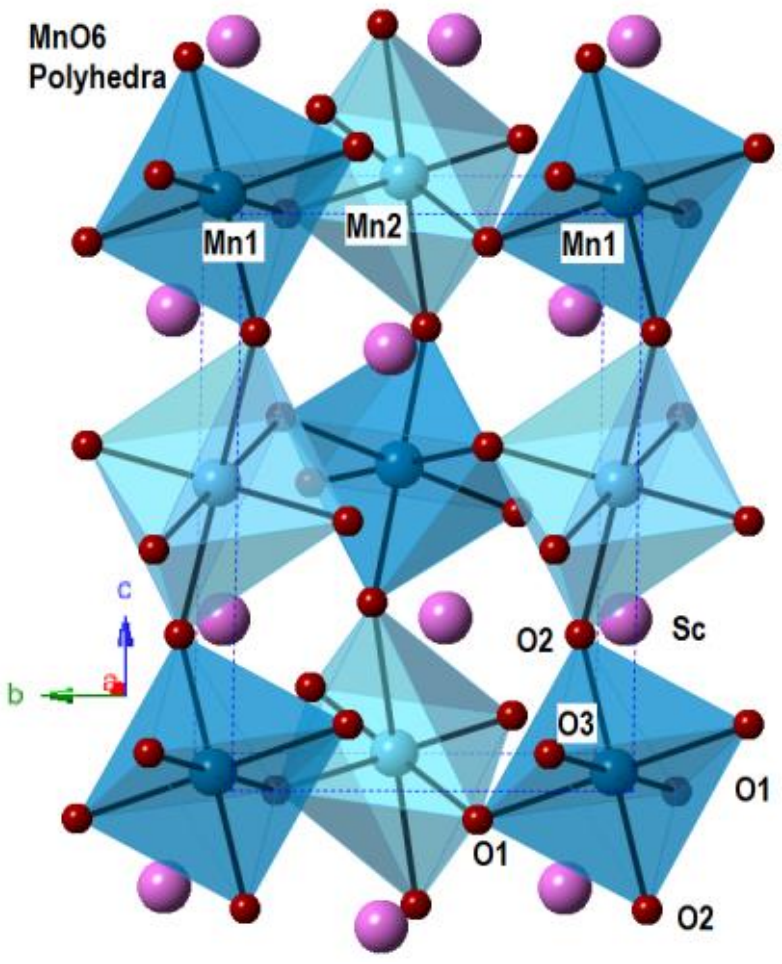

FIG. 1. The crystal structure of monoclinic $ScMnO_3$. The space group of $ScMnO_3$ is $P2_1/n$ with 4 formula units in each cell. Unit cell given as dot line. The positions of the unique atomic sites are labeled for clarity and the unique $MnO_6$ polyhedral darker (Mn1) and lighter (Mn2) shading. In the figure, large, intermediate and small spheres correspond to Sc, Mn and O, respectively.

DFT simulation of the magnetically ordered states of different spin configurations were conducted for the $LuMnO_3$ and $ScMnO_3$ system. The level diagrams for the $LuMnO_3$ and $ScMnO_3$ systems in Figure 2 shows the relative energy for the ferromagnetic (FM), A-type antiferromagnetic (A-AFM), E*-type antiferromagnetic (E*-AFM) and E-type antiferromagnetic states. In addition, the total electric polarization $P_i$ in the E-AFM state relative to the A-AFM state is given. The polarization values typical of rare earth systems predicted in the previous work[6] were found for $LuMnO_3$ with the predicted E-phase as the lowest energy magnetic phase. $LuMnO_3$ is also predicted to move from the centrosymmetric Pnma space group at high temperature (modeled by the A-AFM state) to the lower symmetry polar space group $Pmn2_1$ in the magnetically order low temperature state (E-AFM). For the $ScMnO_3$ system the high temperature structure conforms to the centrosymmetric $P2_1/n$ space group while the low temperature phase transforms to the polar Pc space group for E-phase magnetic ordering at low temperature. Figure 3 shows the structure of the zig-zag spin configuration of the E-AFM phase and the displacement vectors for the Mn and O sites for the $LuMnO_3$ and $ScMnO_3$ systems, predicted by our DFT calculations. The vectors in Figure 3(b) are drawn in proportion to the atomic displacements. The DFT model predicts that the Mn ions will show the largest displacement from the high symmetry positions with the displacement in $ScMnO_3$ two times larger (~0.073 Å) than those of the $LuMnO_3$ system (~0.035 Å). Hence, if there is a large distortion in the E-phase systems, $ScMnO_3$ is an excellent candidate in which to search for them at low temperature. Consequently, we carried out detailed temperature- dependent structural measurements on this system down to 10 K. In addition, the $LuMnO_3$ system was studied with XAFS spectroscopy and PDF measurements. Note that with the low contrast (similar Z for all atoms) of electronic charge for $ScMnO_3$, compared with, for example $LuMnO_3$, X-ray scattering methods will yield accurate assessment of the behavior of the O sites in addition to the Mn and Sc sites, as the temperature is varied in the $ScMnO_3$ system. The same scattering measurements would allow the probing of the mainly heavy Lu and Mn sites by x-rays in $LuMnO_3$.

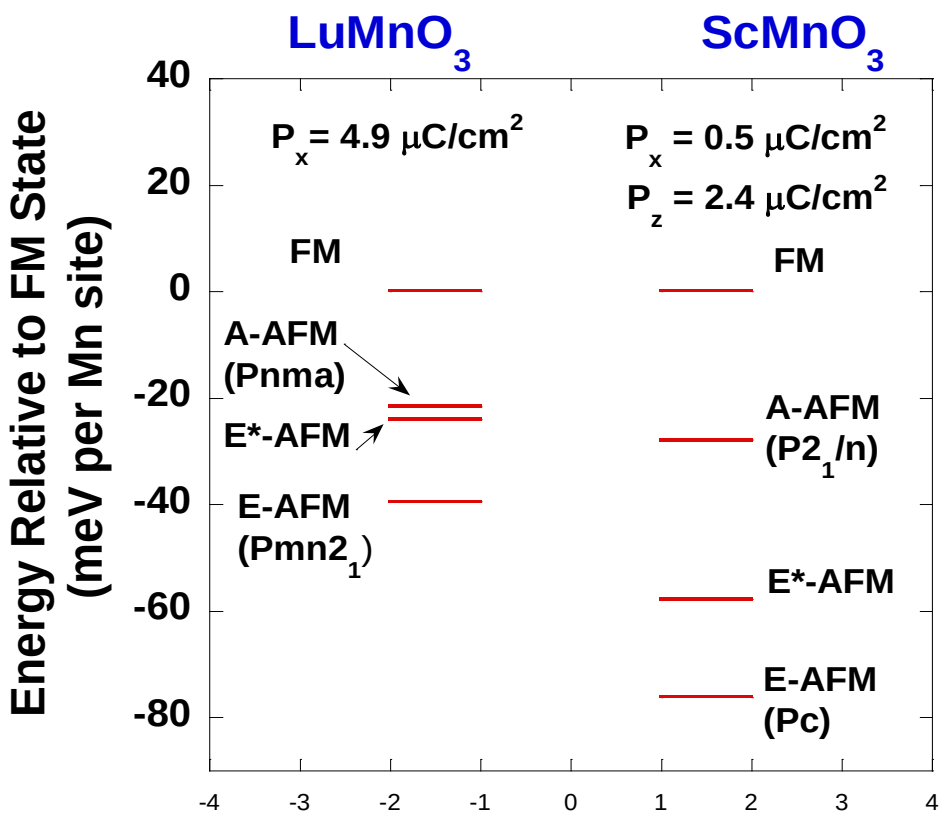

FIG. 2. DFT derived energy level diagram for perovskite $LuMnO_3$ and $ScMnO_3$ from the simulations indicating that the lowest energy magnetic phase is the E-AFM phase with reduced symmetry for both systems ($Pmn2_1$ for $LuMnO_3$ and Pc for $ScMnO_3$).

* E-mail: : tyson@njit.edu

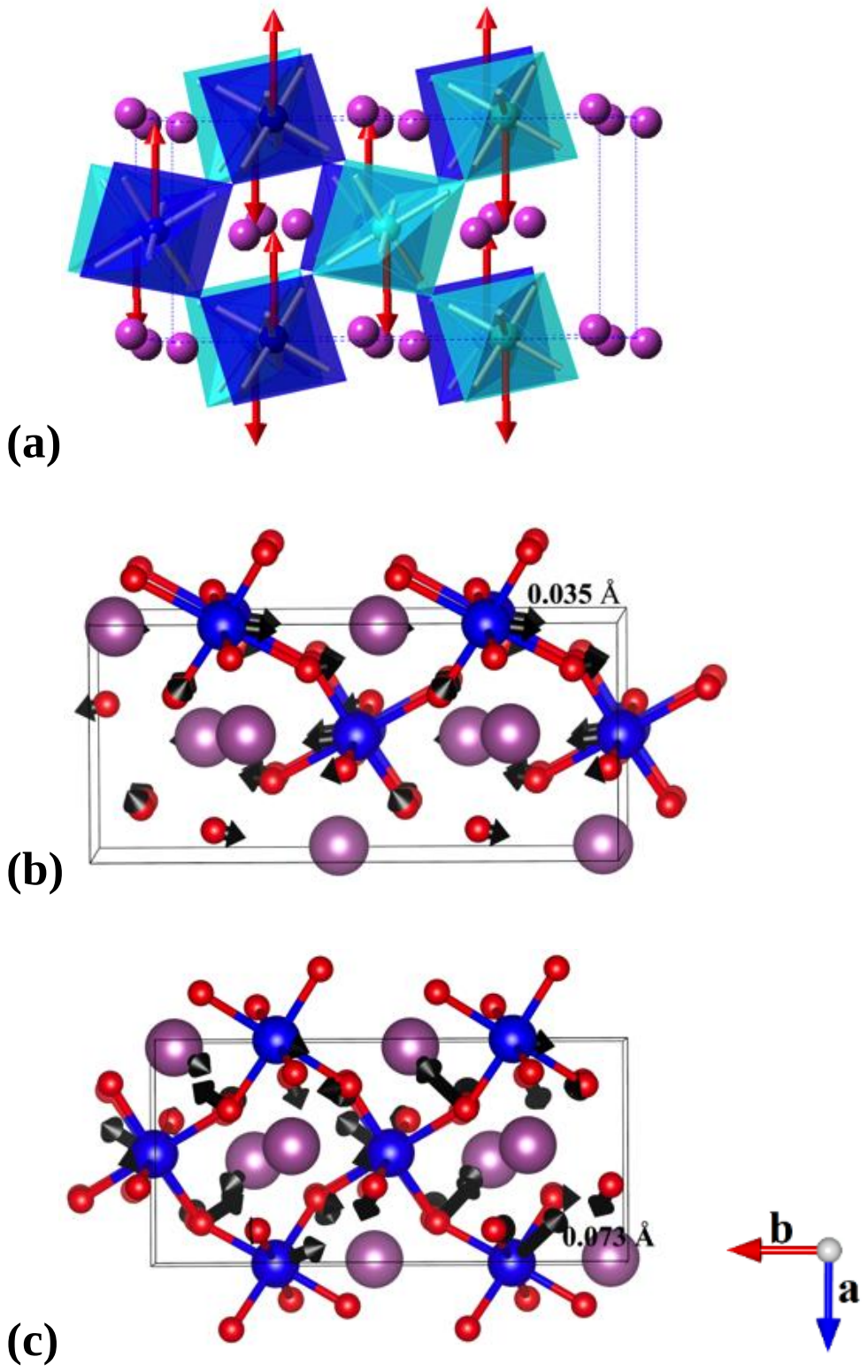

FIG. 3. The magnetic order in the E-phase is given in (a) and the displacement of atoms relative to the A-phase magnetic order is given in (b) for $LuMnO_3$ and (c) for $ScMnO_3$. Note the predicted displacements are twice as large for the $ScMnO_3$ system compared to $LuMnO_3$.

### B. Magnetic Order

Heat capacity measurements reported in Figure 4 reveal the same behavior at low temperature as seen in other E-phase perovskites[21] with rare earth ions except the transition is ~10 K higher ($T_N$ ~ 51 K). Zero-field-cooled (ZFC, cooled from 300 K to 2 in zero field and then warmed in 1kOe, 20 kOe or 50 kOe magnetic field) and field cooled (FC, cooled from 300 K to 2 in 0.1 T, 2 T and 5 T) measurements of the magnetization were conducted. The results reflect the magnetic nature of the transition which onsets near the ~51 K peak found in the heat capacity measurement as shown in Figure 4(b). At low fields (0.1000 T), the shapes of the curves are similar to those in spin glasses[20] or in systems with weak ferromagnetism accompanying an antiferromagnetic state with strong magnetic anisotropy[21]. However, the frequency-dependent magnetic susceptibility measurements show no shift, ruling out the possibility of a spin glass state. Extrapolation of the inverse susceptibility using the Curie-Weiss law for the paramagnetic phase (using data between 100 and 350 K) for the 2 T data yielded a value of the Curie-Weiss temperature of $\theta_p$ = -100 K as shown in Figure 4(b). This value should be compared to values of -62 K to -100 K found for orthorhombic $LuMnO_3$[22]. Not that the ratio f = $-\theta_p/Tc$, so-called frustration parameter[23], is near 3 for simple fcc antiferromagnets like MnO and attains values > 10 for strongly frustrated system such as spin on a triangular lattice. The value f ~2 for $ScMnO_3$ is slightly lower than f ~ 2.5 for the standard orthorhombic system $LuMnO_3$. Hence $ScMnO_3$ exhibits AFM-type interactions, which follow those of the standard orthorhombic E-phase systems such as $LuMnO_3$ but with a lower level of spin frustrations. Detailed structural studies were conducted in order to follow any local or long-range atomic displacements concomitant with the onset of the E-AFM phase. It is noted that both $ScMnO_3$ and $LuMnO_3$ have no spins at the R site (as in the case of for example $HoMnO_3$) making these systems suitable for testing existing models of the E-phase.

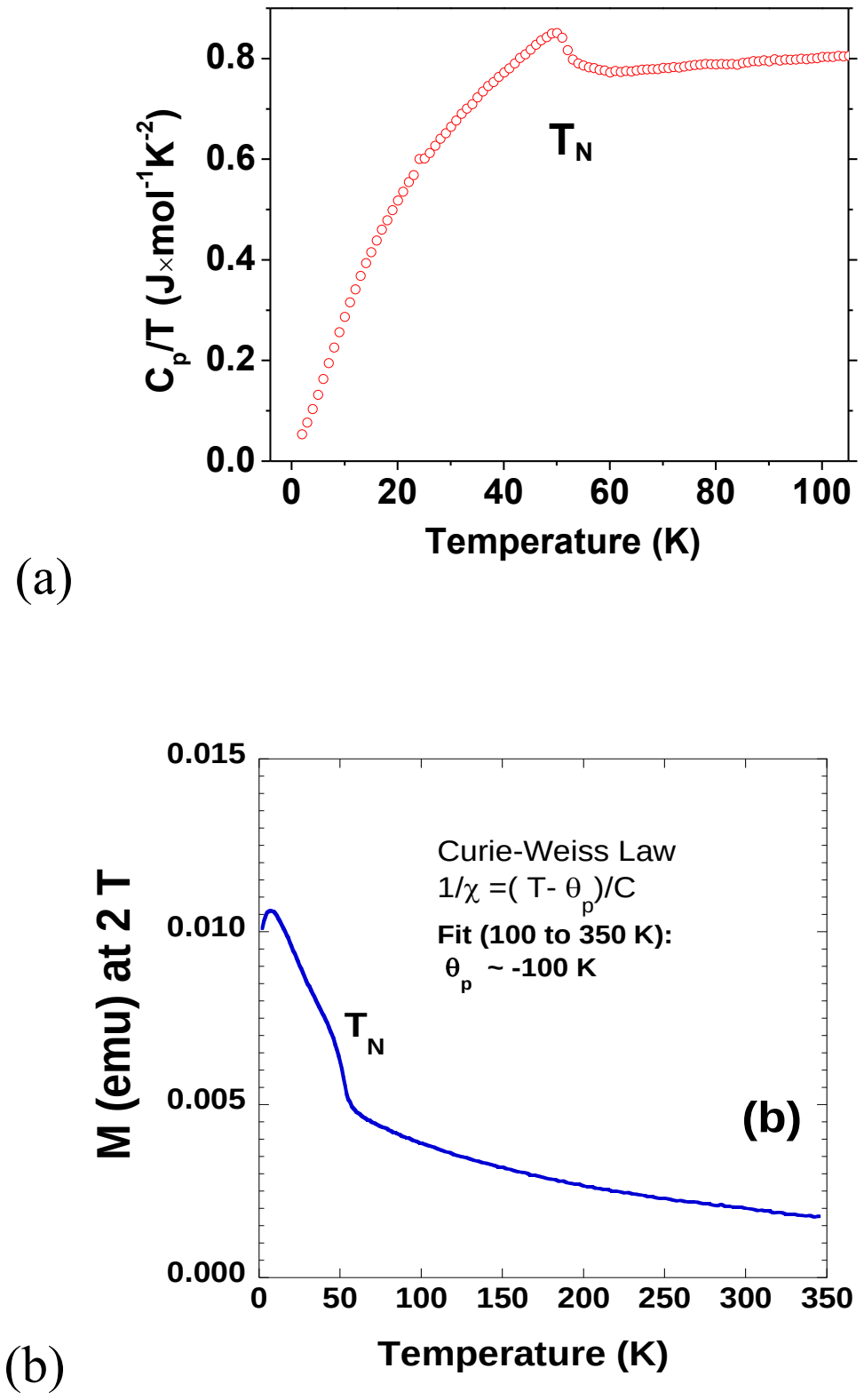

FIG. 4. (a) Heat capacity and (b) magnetization curves indicate that the transition near ~51 K is the AFM ordering temperature ($T_N$) with Curie Weiss temperature ~ 100 K.

### C. Structural Measurements

X-ray pair distribution function (PDF) measurements (10 to 300 K in temperature) were used to explore magnetically induced changes in structure in $ScMnO_3$. The high temperature $P2_1/n$ structure was used as a model for all temperatures. In Figure 5, it is seen that $R_w$

$$R_W = \frac{\sum_{i=1}^{N} w(r_i)[G_{Obs}(r_i) - G_{Calc}(r_i)]^2}{\sum_{i=1}^{N} w(r_i)[G_{Obs}(r_i)]^2} \qquad (1)$$

where $G_{Obs}$ and $G_{Calc}$ are the observed and calculated PDFs and $w$ is the weighting factor; $w(r_i) = 1/\sigma^2(r_i)$, where $\sigma$ is the estimated standard deviation on the data-point at position $r_i$[15(a)], the Goodness-of-fit parameter, varies continuously with temperature. Note that G(r) is the reduced atomic pair distribution function which oscillates about zero and is obtained directly from the scattering data, S(Q). The function

$$G(r) = \frac{2}{\pi}\int_0^{\infty} Q[S(Q)-1]sin(Qr)dQ \qquad (2)$$

is related directly to the standard pair distribution function $g(r)$. ( $G(r) = 4\pi r\rho_0[g(r)-1]$ where $\rho_0$ is the number density of atoms.) As temperature is reduced it is seen that the fit to the high temperature phase $P2_1/n$ structure gets better, contradicting the DFT results which predicts a transition to a lower symmetry Pc space group. The overall scale factor for the data is also shown indicating that there is no observed anomaly as the temperature is lowered. Fit over two r-space region $r_{max}$ = 15 Å (short range) and $r_{max}$ = 40 Å (intermediate range) yield identical trends, see Figure 5.

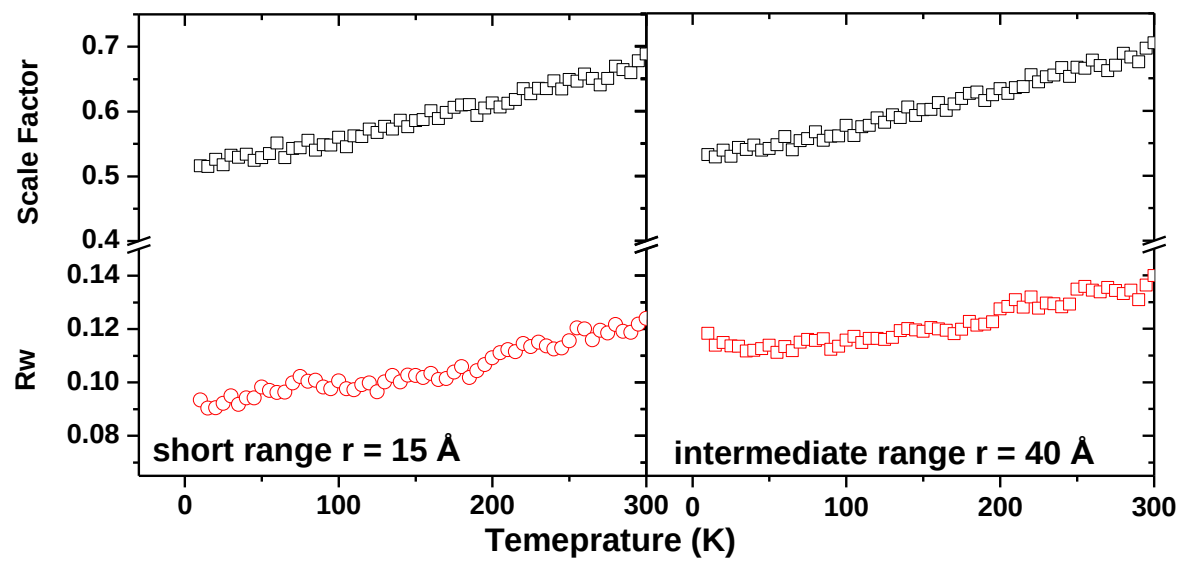

FIG. 5. The scale factor and Rw fitting parameters after refinement of PDF experimental data, the left set gives the result for short range fitting with r up to 15 Å, the right set is the result for intermediate range fitting with r up to 40 Å. Note that in all cases the fits improve as temperature is decreased.

Examining the details the reduced pair distribution function makes the nature of structural changes near $T_N$ more clear (Figure 6). The low-R peaks (below ~5 Å) show no anomalies on crossing $T_N$. The normal broadening of the Mn-O, Sc-O, or Mn-Mn peaks is seen in Figure 6 as temperature is increased. Figures 6(c) and 6(d) show that fits over a shorter range give better fits indicating the presence of significant local temperature independent structural distortions.

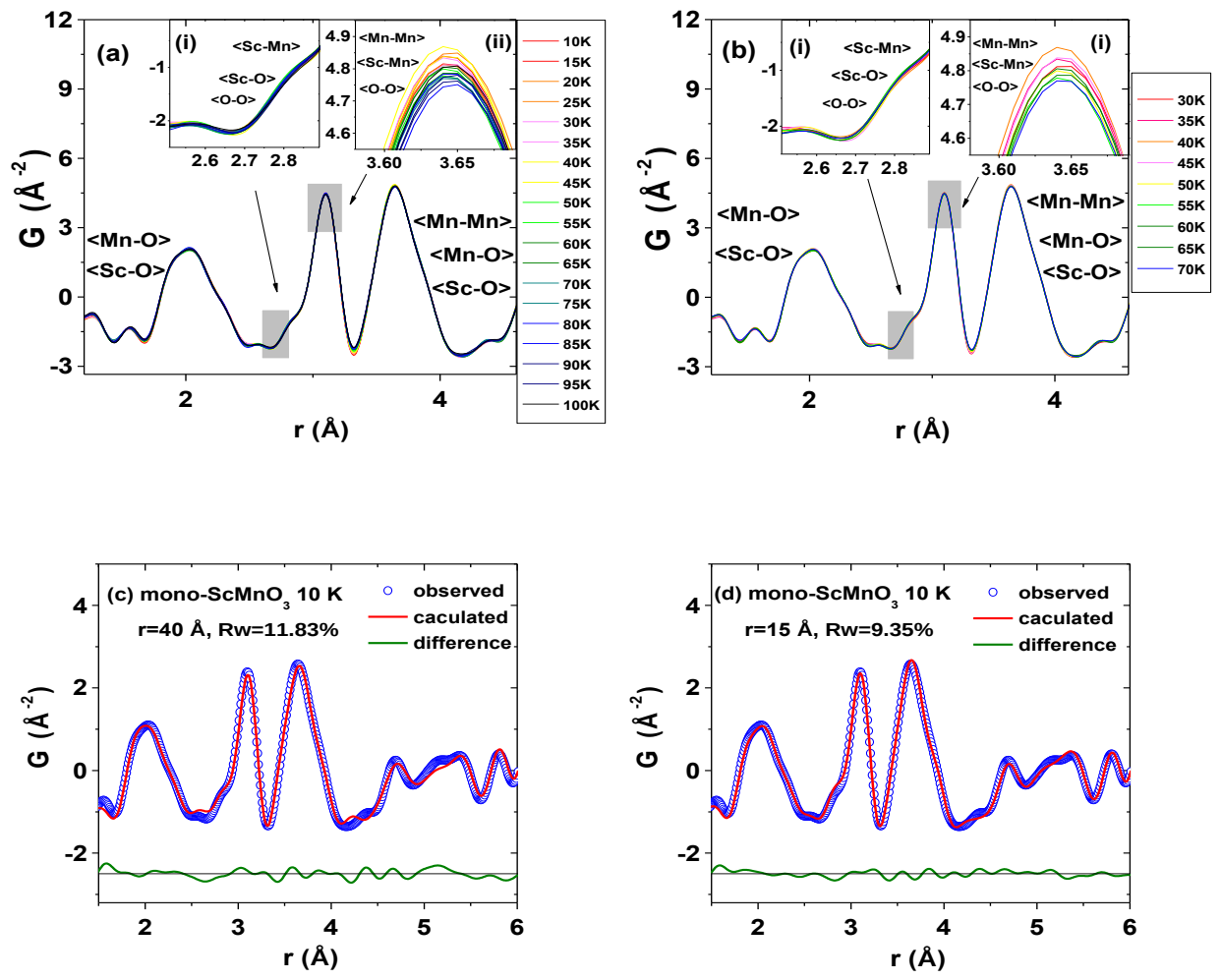

FIG. 6. (a) The temperature dependence of the local structure for PDFs (10 to 100 K). The bond distances from the refinement are used to label the peaks. (b) The temperature dependence of the local structure near the Néel temperature (30 to 70 K). Insert (i) is the expanded range for O-O (at 2.47~2.57 Å), Sc-O (at ~2.81 Å) and Sc-Mn (at ~2.89 Å) peaks. (ii) is the expanded range for O-O (at ~3.62 Å), Sc-Mn (at ~3.63 Å) and Mn-Mn (at ~3.68 Å) peaks. (c) A comparison of the $P2_1/n$ model and the observed PDF data at 10 K with an intermediate range ($r_{max}$ = 40 Å) fitting. (d) A comparison of the $P2_1/n$ model and the observed PDF data at 10 K with short range ($r_{max}$ = 15 Å) fitting. The improvement with reduce fitting range indicates the presence of temperature independent local distortions.

Similarly, Figure 7 reveals no anomalies of the lattice parameters at $T_N$. It is found that the volume and angle, β exhibit no abrupt changes at low temperature either. Examination of the isotropic atomic displacement parameters (ADPs) of Mn also reveals no change near $T_N$, as shown in Figure 8. As mentioned above, fits with $r_{max}$ = 5 Å, 15 Å and 40 Å reveal that a feature near 5 Å is not modeled well by the room temperate space group at low temperature but the local distortion starts to appear at temperatures significantly above $T_N$.

* E-mail: : tyson@njit.edu

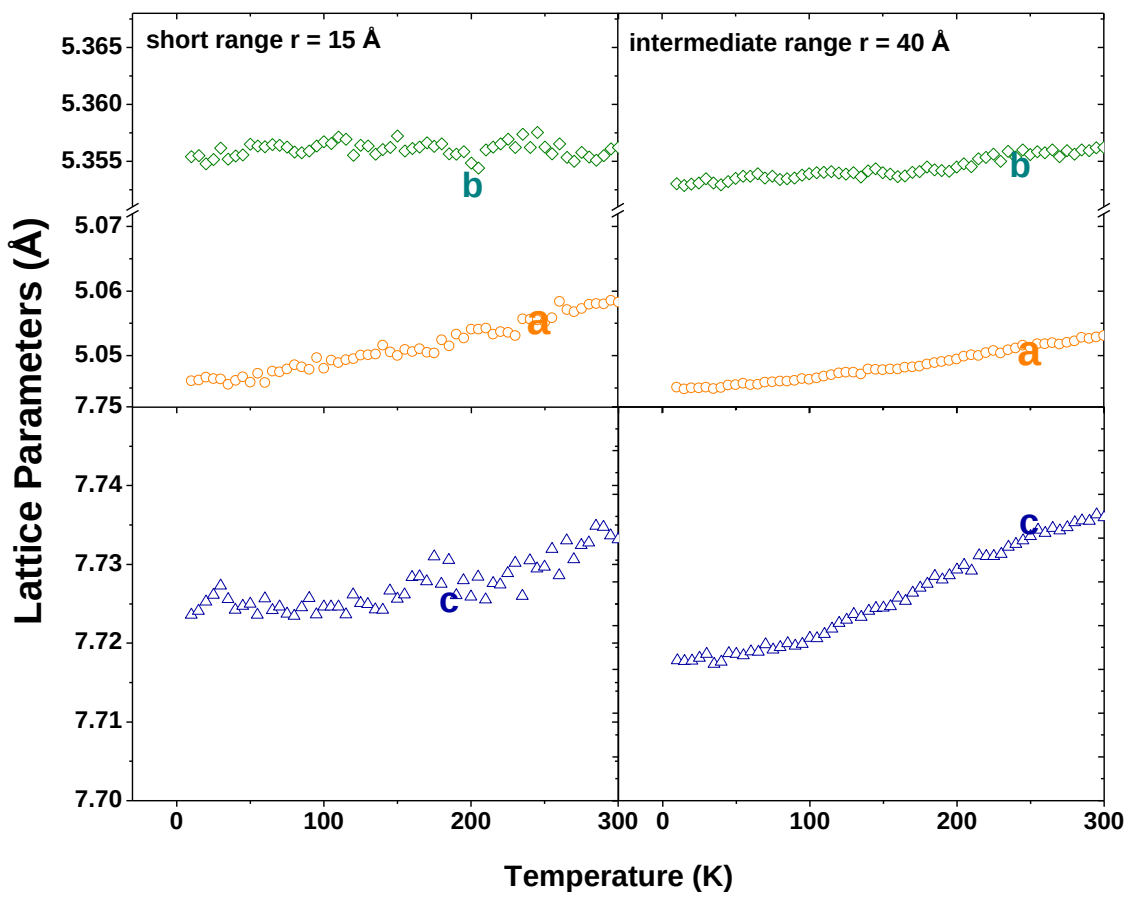

FIG. 7. The temperature dependent lattice parameters, a, b and c are compared both in short range (r = 15Å) and intermediate range (r = 40Å).

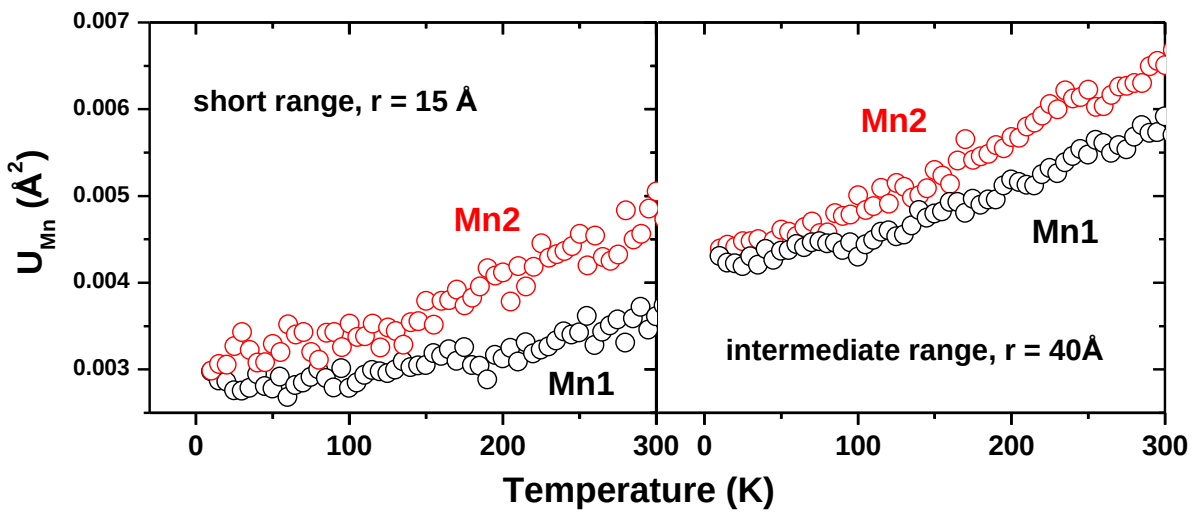

FIG. 8. The atomic displacement parameters (ADPs, isotropic) for Mn1 and Mn2 are shown through the temperature range of 10 K to 300 K, both in short range ($r_{max}$ = 15Å) and intermediate range ($r_{max}$ = 40Å) fits. There is no obvious change over this temperature range.

To probe the changes in long-range structure more precisely, single crystal X-ray diffraction measurements were conducted in the temperature range of 10 K to 60 K. Figure 9 shows the R1 fitting parameter defined as,

$$R_1 = \frac{\sum_{i=1}^{N} ||F_{Obs}|_i - |F_{Calc}|_i}{\sum_{i=1}^{N} |F_{Obs}|_i} \quad (3)$$

where $F_{Obs}$ is measured structure factor and $F_{Cal}$ is calculated structure factor, for the given temperature range around $T_N$. No systematic trend is seen except for statistical errors. The lattice parameters a, b, c and β (Figure.10) also show no statistically significant change. It is found that β increases as temperature is reduced. To study the local structural anomalies, the ADPs were also examined. It is seen that the anisotropic ADPs ($U_{ij}$) of all sites (Figure 11) show no anomalies near $T_N$. In fact, the anisotropic ADPs suggest a trend to higher symmetry (reduction in amplitude with lowering of temperature), consistent with the $R_w$ parameters from PDF measurements (see Figure 5). No significant changes are seen in the Sc-Mn bond distances (Figure 12 and Table I).

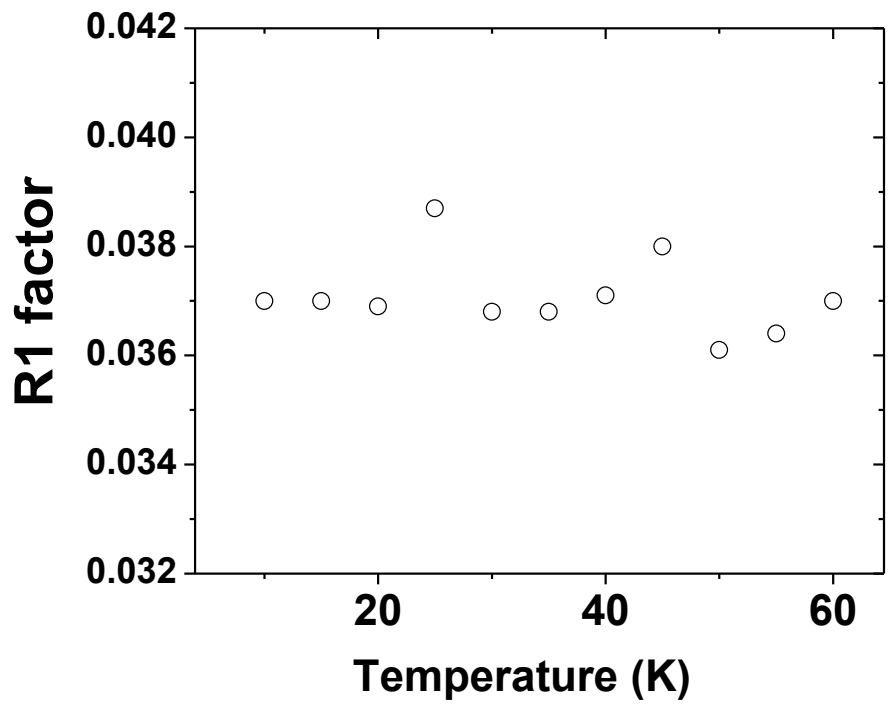

FIG. 9. The R1 factor after refinement of single crystal XRD experimental data.

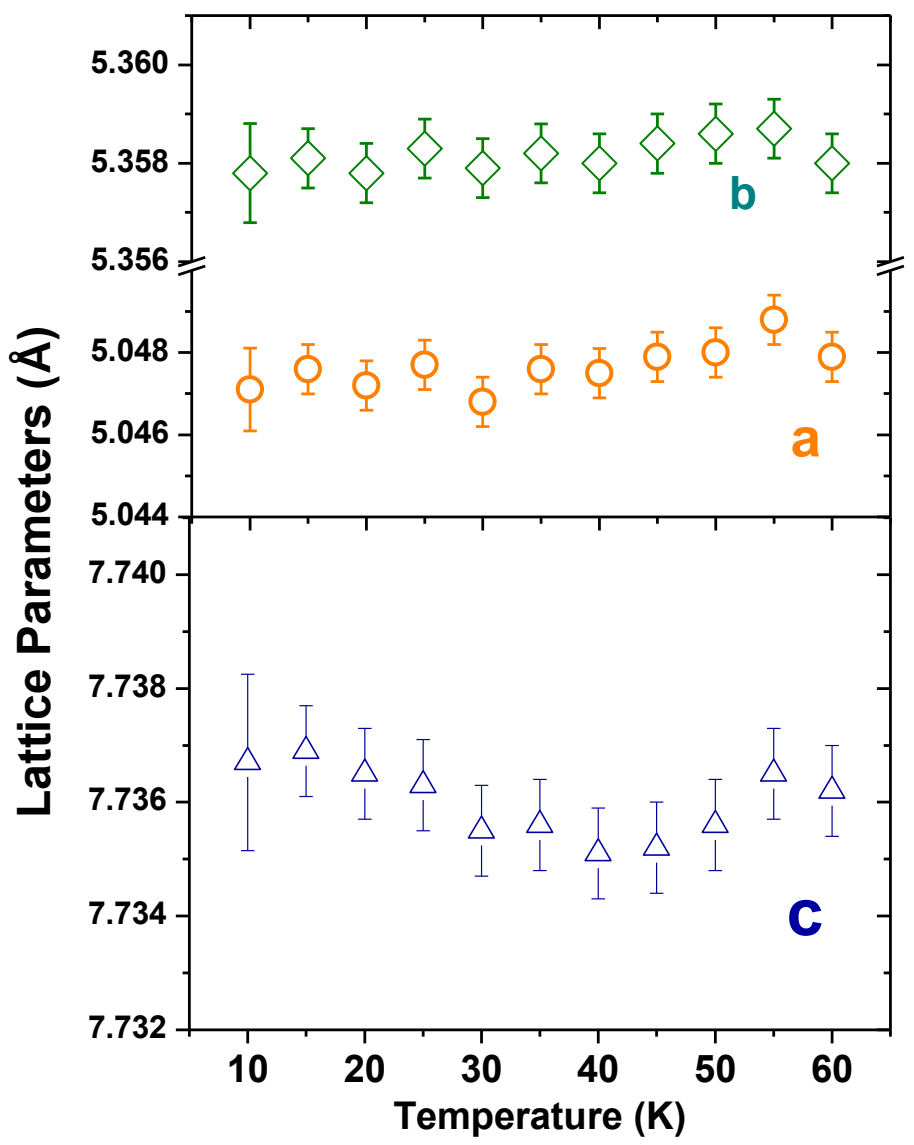

FIG. 10. The lattice parameters, a, b and c are compared with temperature showing that there is no obvious change in this temperature range which includes $T_N$.

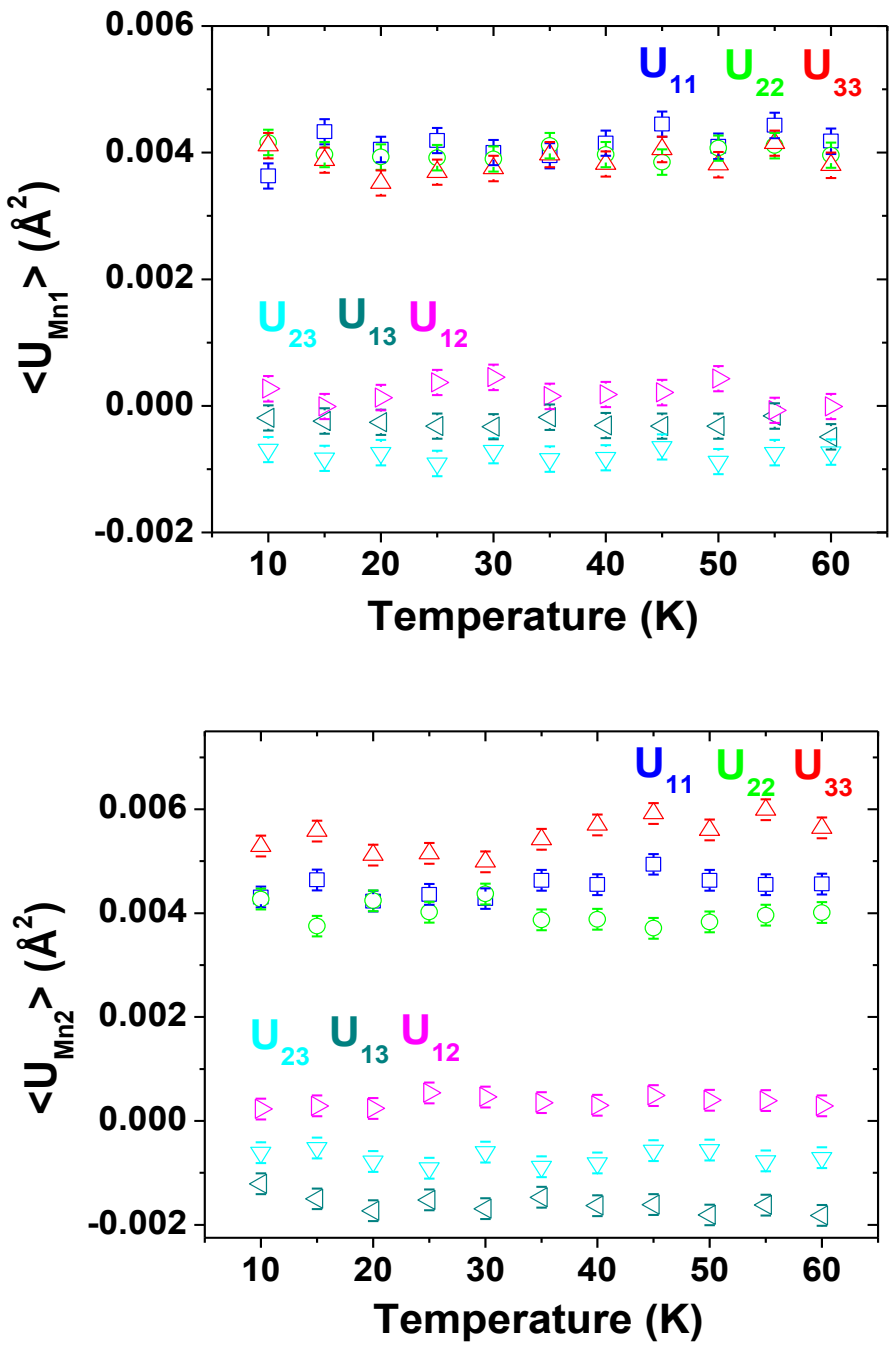

FIG. 11. The temperature dependent atomic displacement parameters (ADPs) for Mn1 and Mn2. There is no obvious change in the temperature range of 10 to 60 K.

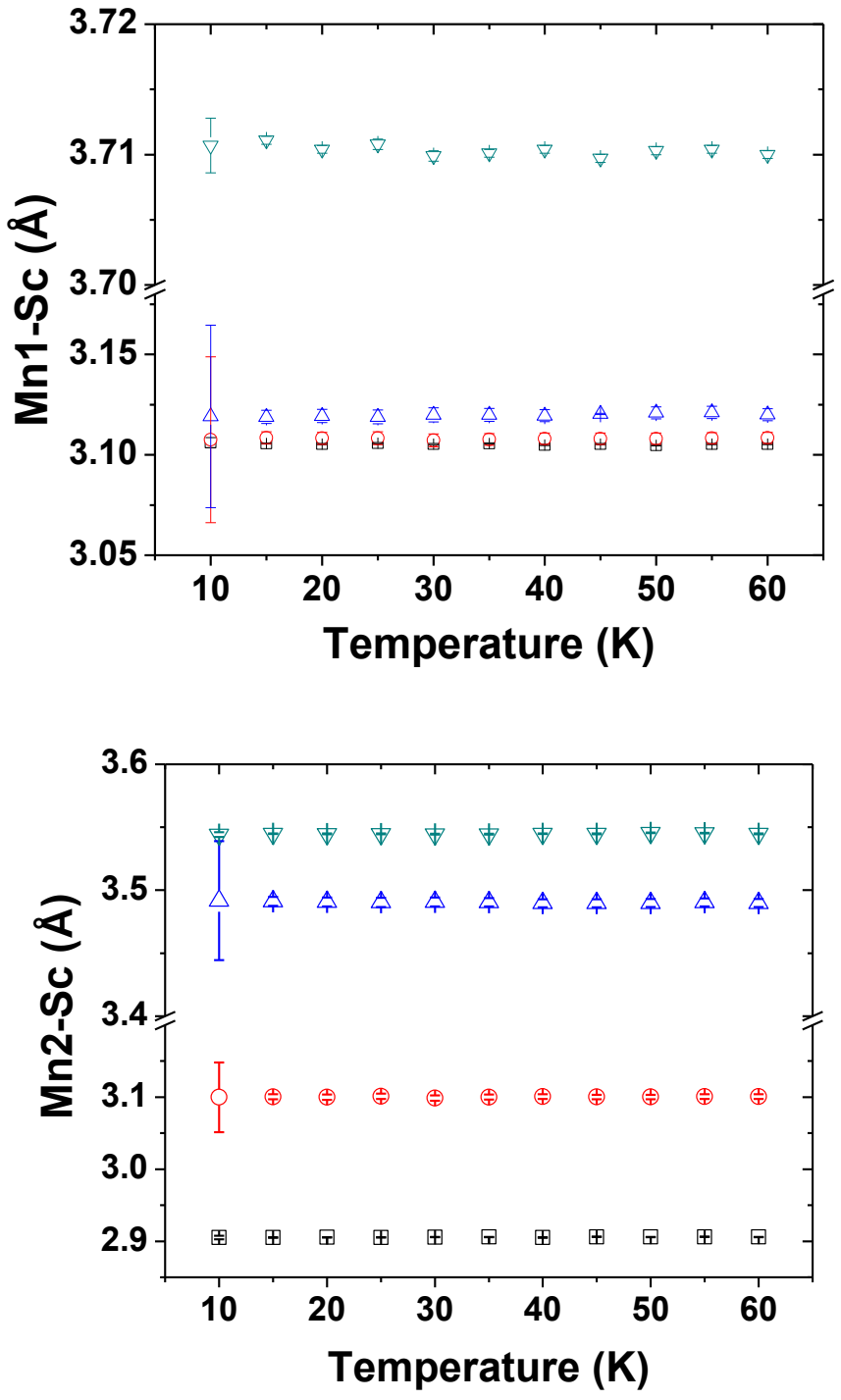

FIG. 12. The temperature dependent bond distances of Mn-Sc. There is no obvious change in the temperature range of 10 to 60 K.

**Table I.** Changes in Perovskite $ScMnO_3$ and $LuMnO_3$ Structural Parameters with Temperature (High Temperature Minus Low Temperature).

| Single Crystal XRD | | PDF | | XAFS | |
|---|---|---|---|---|---|
| Δa (Å) | 0.00100(51) | Δa (Å) | 0.00063(17) | $\Delta R_{Sc\text{-}Mn}$ (Å) | -0.0015(38) |
| Δb (Å) | 0.00400(38) | Δb (Å) | 0.00073(8) | $\Delta\sigma^2_{Sc\text{-}Mn}$ ($Å^2$) | -0.0001(20) |
| Δc (Å) | -0.00045(18) | Δc (Å) | 0.00080(22) | | |
| Δβ (deg.) | -0.0303(34) | Δβ (deg.) | 0.0154(38) | **$LuMnO_3$** | **(Lu-Mn)** |
| $\Delta R_{Sc\text{-}Mn}$ (Å) | -0.0015(38) | $\Delta<U_{Mn1}>$ ($Å^2$) | 0.00014(5) | $\Delta R_{Lu\text{-}Mn}$ (Å) | 0.00085(570) |
| $\Delta\sigma^2_{Sc\text{-}Mn}$ ($Å^2$) | -0.0001(20) | $\Delta<U_{Mn2}>$ ($Å^2$) | 0.00015(4) | $\Delta\sigma^2_{Lu\text{-}Mn}$ ($Å^2$) | 0.00023(11) |
| $\Delta<U_{Sc}>$ ($Å^2$) | 0.00019(10) | $\Delta<U_{Sc}>$ ($Å^2$) | 0.00015(5) | $\Delta R_{Lu\text{-}Mn}$ (Å) | 0.00075(470) |
| $\Delta<U_{O1}>$ ($Å^2$) | 0.00009(21) | $\Delta<U_{O1}>$ ($Å^2$) | 0.00046(21) | $\Delta\sigma^2_{Lu\text{-}Mn}$ ($Å^2$) | 0.00014(24) |
| $\Delta<U_{O2}>$ ($Å^2$) | -0.00002(22) | $\Delta<U_{O2}>$ ($Å^2$) | -0.00045(9) | $\Delta R_{Lu\text{-}Mn}$ (Å) | -0.0013(9) |
| $\Delta<U_{O3}>$ ($Å^2$) | 0.00024(6) | $\Delta<U_{O3}>$ ($Å^2$) | 0.00007(35) | $\Delta\sigma^2_{Lu\text{-}Mn}$ ($Å^2$) | 0.00015(30) |

Note: The differences between lattice parameters and atomic displacement parameters around Néel temperature are listed. For single crystal XRD and PDF data, below $T_N$ the data used are 10 K and 15 K, above $T_N$ are 55 K and 60 K. For XAFS, data below $T_N$ are at 16 K and 30 K, above $T_N$ are at 60K and 70K for $ScMnO_3$. For the $LuMnO_3$ system the 10 K and 15K data were compared to 50 K and 60 K. The parameters such as $<U_{Mn1}>$ correspond to the average of the diagonal terms of the $U_{ij}$ matrix.

Changes within the $MnO_6$ polyhedra can be characterized by utilizing the Jahn-Teller distortion parameters[24],

$$Q_2 = (l - s)\sqrt{2} \tag{4}$$

and

$$Q_3 = (2m - l - s)\sqrt{\frac{2}{3}} \tag{5}$$

where $l$ = long bond distance, $m$ = medium bond distance and $s$ = short bond distance, characterize the Mn-O splittings in the $MnO_6$ polyhedra, and are shown in Figure 13 for the temperature range studied. No statistically significant changes in these parameters are seen upon magnetic. No changes are seen in any bond distances either. In Figure 13, the Q parameters for the E-phase of the $ScMnO_3$ system from the DFT simulation are included as solid symbols showing that the distortions are similar in magnitude to the experimental values. $Q_{2,avg}$ ~ 0.45 Å corresponds to the planar Jahn-Teller distortion $e_{JT}$ ~ 0.0795 defined in Ref. [25], and is consistent with the trend of the increased Jahn-Teller distortion for $RMnO_3$ with smaller rare earth ions due to the coupling among chemical potential, shear distortion, and Jahn-Teller distortion studied in Ref. [24].

* E-mail: : tyson@njit.edu

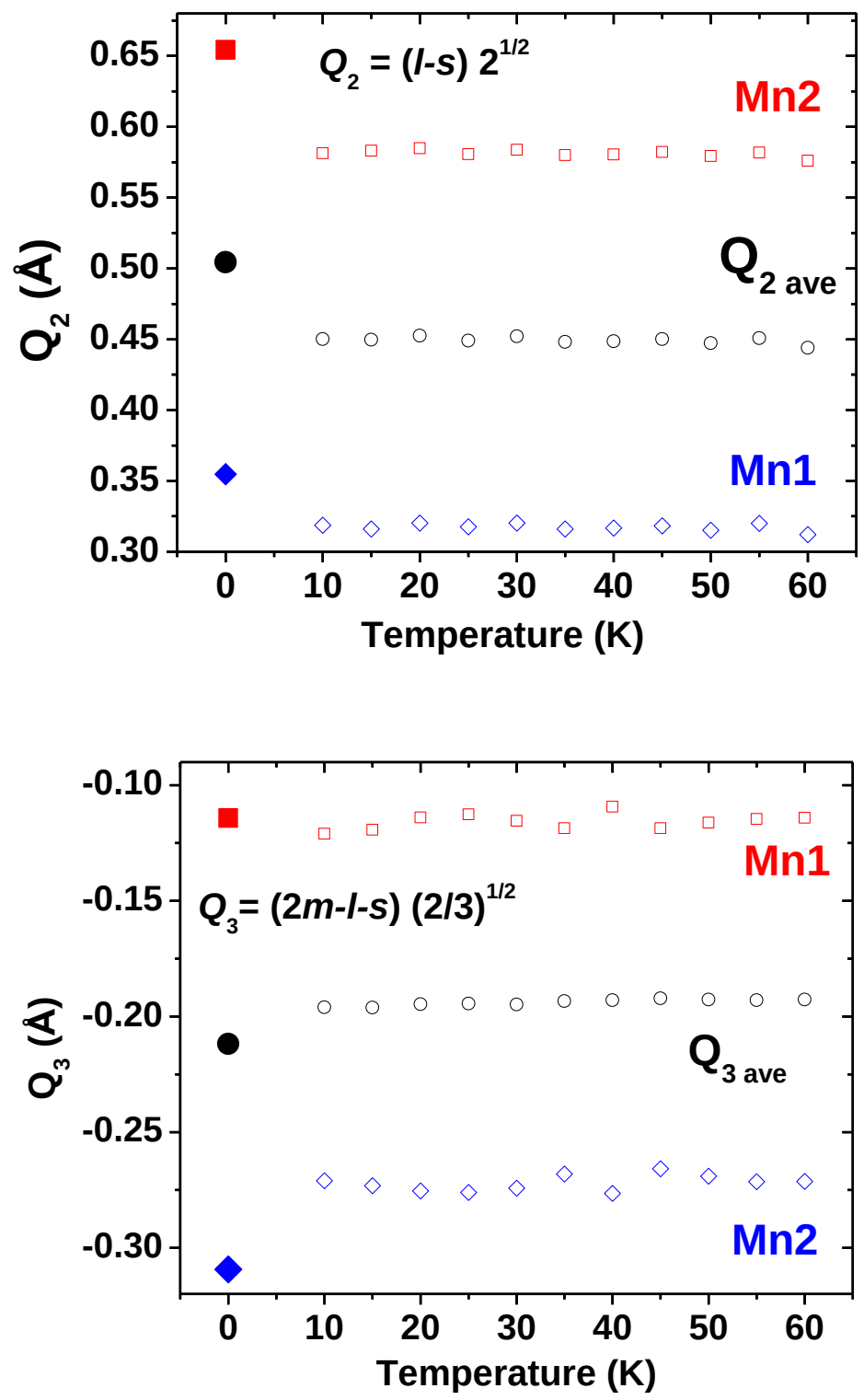

FIG. 13. $Q_2$ and $Q_3$ parameter which characterize the Jahn-Teller distortion of bond distances. Solid symbols are values (at 0 K) from the DFT simulations. No change is seen with temperature.

For $ScMnO_3$, the PDF and single crystal measurements are complemented by the XAFS local structure about Sc and Mn, as shown in Figure 14. The results show no change in the temperature range of 13 K to 75 K. The changes in the structure crossing over to the magnetically ordered E-phase are below the detection limit. Fits of the structure to perovskite model were used to extract atomic level structural details and the difference between the high and low temperature structural paramters. These results are displayed in Table I from all of the conducted measurements. In the case of perovskite $ScMnO_3$, the upper limit on the lattice parameter change is $4 \times 10^{-3}$ Å (single crystal XRD) while that for the change in the Sc-Mn bond (XAFS) is $2 \times 10^{-3}$ Å with no detectable change in width. No evidence for local distortions is found.

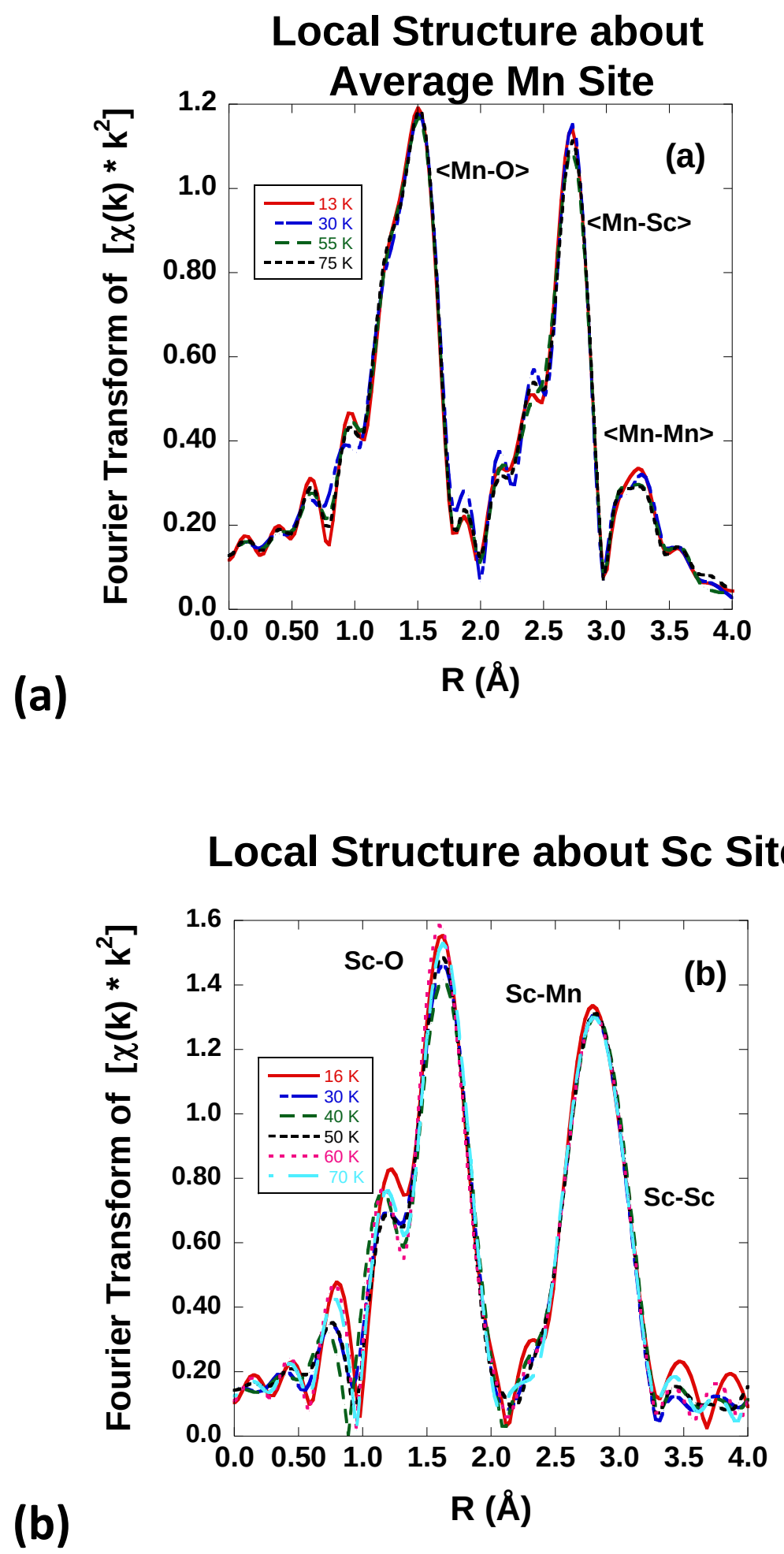

FIG. 14. XAFS structure functions of perovskite $ScMnO_3$ about the average Mn Site (a) and about the Sc site (b) for temperatures above and below the magnetic ordering temperature. Major peak components are labeled. Fits to the Sc K-Edge data Sc- Mn Peak indicate that any shift in the Mn sites is less than $2 \times 10^{-3}$ Å (see Table I).

In terms of the structural changes for the $LuMnO_3$ system, neutron powder diffraction measurements at 8 K and 298 K on this system have previously been conducted[20] and structural refinements were conducted within the Pbnm space group. No statistically significant change in the isotropic ADP values of Mn is found for these temperatures. Interestingly, the ADPs of oxygen are larger at 8 K than at 298 K. These results again contradict the DFT model. We extended this work with PDF and XAFS measurements. In the X-ray PDF data on $LuMnO_3$ in the temperature range of 11 to 65 K (Figure 15), the Mn-O and Lu-O peaks are identified. Two of the peaks are dominated by the Lu-Mn bonds. The change in the Mn-O and Lu-O peaks observed here are artifacts due to the weak scattering of x-rays by O atoms compared to Mn and Lu. However, the stronger peaks dominated by Lu-Mn distance near ~3 Å exhibit sharpening as temperature is reduced, instead of the expected broadening due to lowering in symmetry as predicted by DFT models. XAFS measurements of the Lu site, probing the Lu-O, Lu-Mn and Lu-Lu bond distributions, were conducted. The structure functions about the Lu site are shown in Figure 16. All peaks including the Lu-Mn peaks are enhanced at low temperature indicating no distortion enhancement with reduced temperature. Fit to the data were used to quantify the Lu-Mn peak positions which are split into three distinct components. From the XAFS results on $LuMnO_3$, upper limits on the $LuMnO_3$ displacements are $1 \times 10^{-3}$ Å (See Table I).

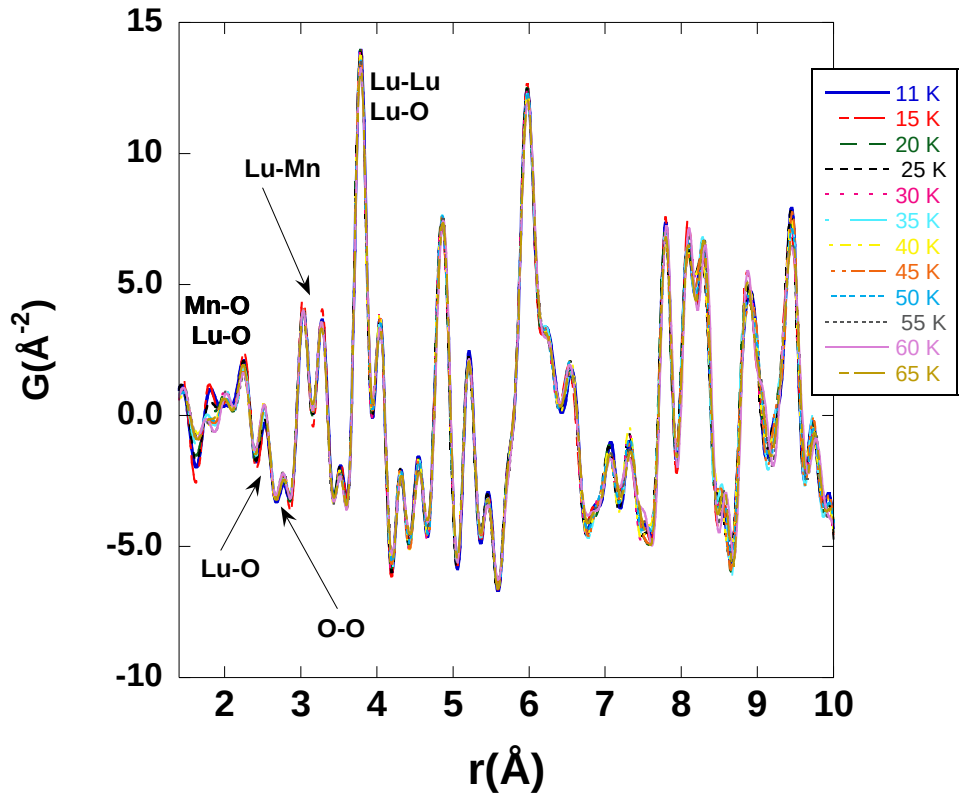

FIG. 15. PDF reduced distribution function of perovskite $LuMnO_3$ are shown for data taken between 11 and 65 K. Examining the Lu-Mn peaks one sees now significant changes (broadening or shifts).

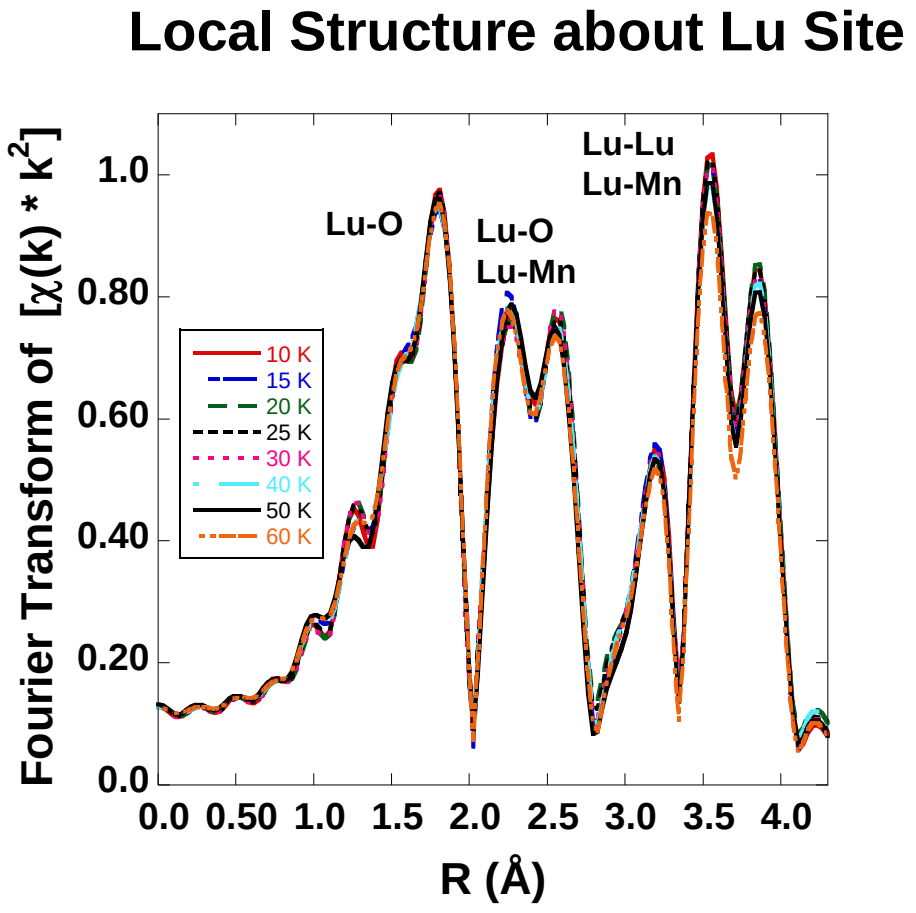

FIG. 16. XAFS structure functions of perovskite $LuMnO_3$ about the average Lu site. Fits to the Lu L3-Edge data Lu-Mn peaks indicate that any shift in the Mn sites is less than $1 \times 10^{-3}$ Å (see Table I).

## IV. SUMMARY

In summary, the detailed structural measurements reported in this paper reveal no significant local or long range distortions associated with the E-phase of $ScMnO_3$ and $LuMnO_3$ at a level greater than $\sim 10^{-3}$ and contradict the prediction of DFT calculations by an order of magnitude. It suggests that standard DFT models of the E-phase system are not adequate and indicate that DFT does not properly account for the spin-lattice coupling in these oxides and possibly it may have predicted the incorrect magnetic order. The results suggest that the electronic contribution to the electrical polarization dominates and should be properly treated in models.

## V. ACKNOWLEDGEMENTS

This work is supported by DOE Grant DE-FG02−07ER46402. Synchrotron powder X-ray diffraction and X-ray absorption data acquisition were performed at Brookhaven National Laboratory's National Synchrotron Light Source (NSLS) which is funded by the U.S. Department of Energy. Single crystal X-ray diffraction measurements were performed at ChemMatCARS Sector 15, which is principally supported by the National Science Foundation/Department of Energy under Grant NSF/CHE-0822838. Use of the Advanced Photon Source was supported by the U.S. department of Energy, Office of Science, Office of Basic Energy Sciences, under Contract No. DE-AC02-06CH11357. The Physical Properties Measurements System was acquired under NSF MRI Grant DMR-0923032 (ARRA award).

* E-mail: : tyson@njit.edu